# Reinforcement Learning for Target Jerk Estimation in Interceptor Homing Phase

Brian Gaudet*
*University of Arizona, 1127 E. Roger Way, Tucson Arizona, 85721*

Dalton Durant†
*University of Arizona, North Cherry Avenue, Tucson Arizona, 85721*

Roberto Furfaro‡
*University of Arizona, 1127 E. Roger Way, Tucson Arizona, 85721*

**We demonstrate the use of reinforcement meta-learning (meta-RL) to optimize a target jerk estimation policy, which is then used as the target acceleration rate of change in the propagation model of an unscented Kalman filter (UKF). The policy is learned through simulated engagements of a terminal phase high-altitude endoatmospheric interception of a highly maneuverable target, with a hit-to-kill accuracy requirement. As benchmarks, we tuned both a standard UKF and an interacting multiple model (IMM) filter, where the IMM uses a different constant jerk assumption for each model. The benchmarks and the proposed augmented UKF with the target jerk policy are tested in a 6-DOF simulation environment, which has sufficient fidelity to capture most factors relevant to guidance, navigation, and flight control. The augmented UKF demonstrates significant improvement in target interception rate, with more challenging maneuvers giving the largest performance differential. The meta-RL jerk policy is then tested on an expanded engagement scenario, with increased initial state estimate uncertainty and novel target maneuvers, demonstrating that the meta-RL policy generalizes well to conditions not seen during optimization.**

## I. Introduction

The next generation of strike weapons fielded by peer adversaries will be capable of actively maneuvering to evade interception through the execution of complex maneuvers. Indeed, rather than executing discrete maneuver modes, targets may continuously vary acceleration magnitude, direction, and frequency in response to interceptor behavior. The capability of a target to execute arbitrary maneuvers during the intercept homing phase creates problems for interceptor guidance, state estimation, and flight control during the homing phase, particularly for the case of an interceptor limited to control surface actuation in a high-altitude intercept scenario against a highly agile hypersonic target with hit-to-kill accuracy requirements. This scenario typically results in short homing times, high body rates, significant radome refraction, and errors in line-of-sight (LOS) recovery, with the last two creating parasitic paths between navigation, guidance, and flight control systems.

In the absence of significant target maneuvers, the apparent target acceleration is governed by slowly varying effects of drag, gravity, and other small perturbations. Therefore, it may be sufficient to assume dynamics models such as near-constant acceleration [1], near-constant jerk [2], or Singer acceleration [3], where the first two assume either target acceleration or jerk are constant over short intervals, and the third assumes autocorrelated acceleration that follows a first-order Gauss-Markov process which decays to zero in the absence of continued maneuvering. These can be used with standard linear filtering algorithms such as the extended Kalman filter (EKF) [4] and unscented Kalman filter (UKF) [5, 6] to accurately estimate the system's relative position and velocity. However, if a maneuver violates the model assumptions, then it can cause the filter to become overconfident and eventually diverge. These simple motion models are usually tuned to accommodate a specific maneuver class, and anything outside the class will require some amount of retuning.

*Research Engineer, Department of Systems and Industrial Engineering, E-mail: briangaudet@arizona.edu
†Postdoc, Space4 Center, E-mail: ddurant@arizona.edu
‡Professor, Department of Systems and Industrial Engineering, Department of Aerospace and Mechanical Engineering. E-mail: robertof@arizona.edu

A solution is to use adaptive filters. These filters are capable of tuning themselves online to adjust for changes in the dynamics not accounted for by the current model. Early work in adaptive filtering actively increases and decreases model uncertainties when target measurements deviate substantially away from the prediction. For example, Refs. [1, 7, 8] each discuss adaptive Kalman filters that detect target maneuvers and modify the process noise covariance. Other methods also include covariance-matching or innovation-matching to adjust either the process noise or measurement noise covariance (or both) so that the innovation statistics align with the filter's innovation covariance [9]. Methods like these are not only capable of producing robust and consistent estimates, preventing filter divergence, but they are also computationally efficient, requiring little extra implementation effort. However, one limitation is that they primarily modify the assumed uncertainty surrounding a fixed dynamics model, meaning they cannot necessarily identify the direction or temporal structure of an arbitrary target maneuver. Another limitation is that these methods rely on adaptation laws that are typically discontinuous. They produce non-smooth transitions and result in abrupt spikes in error estimates when either increasing or decreasing uncertainties. Hence, there is an inherent tradeoff between rapid maneuver detection and noisy or unstable covariance estimates.

A more sophisticated solution to the adaptive filtering problem is to use multiple model (MM) filters. These methods can provide a more explicit treatment of the maneuver uncertainty by operating parallel linear filters, such as an EKF or UKF, and mixing their estimates. Each filter describes a hypotheses about the state of the system, carrying out its own model assumptions, *e.g.* target acceleration dynamics or process noise statistics. In standard MM estimation, the number of models grows combinatorially over time. In Ref. [10] the authors develop a MM adaptive estimator (MMAE) that fixes the number of models to a useful amount, and uses weights as the adaptation mechanism. In other words, the MMAE extracts a weighted average of the estimates provided by each model, where the weights are updated at each time step. They showed that this reduces onboard computation and provided effective estimates for a ballistic-missile intercept scenario. However, one issue with the MMAE is that if a maneuver is sustained for a period of time, then the probability mass can isolate into one of the weights. This means that a single model will have sole influence over the filter and it will be very difficult (and sometimes impossible) for the filter to recover once the target maneuvers again, causing degeneracy. One solution for MMAE filters is to kill off the models with very low weights and replace them with models with higher weights [1]. They can also be subdivided among the existing models into new models that have better performance characteristics, and in the worst case if a model diverges, then it can be reinitialized to have an estimate closer to the current measurement. However, these techniques are ad hoc.

A state-of-the-art MM filter, employed by many interceptor systems, is the interacting MM (IMM) filter from Ref. [11], and later popularized by Ref. [12]. The IMM filter prevents degeneracy by allowing the estimates of the different models to interact with each other after the update, where models with higher probability effect those with less and vise versa. Specifically, model interactions are represented probabilistically by a Markov chain transition matrix, whose elements are the conditional probabilities of each transition disturbance process. After the update, it is used with the weights of the individual models to compute a *mixing* matrix, whose elements are the probabilities of the target switching from one model to another. This mixing matrix is then used to shift the estimated mean and covariance of each model to more closely align with the more dominant ones. Effectively, this enables smooth transitions between maneuver models during the evolution of the target's behavior, which avoids the issues of the early adaptive filters that can have abrupt spikes in their error estimates, and also avoids the issues of the MMAE filter by being less ad hoc and preventing degeneracy. In Ref. [13], the authors show that the IMM filter can be modeled using known limitations of the target's acceleration, where each model is a different acceleration quantity. In Ref. [14], the authors show successful application of an IMM filter to a semi-active missile homing problem using LOS, range, and range-rate measurements, testing three target models: non-maneuvering, mildly maneuvering, and highly maneuvering.

While the IMM has shown to be useful, it is not without its limitations. First, the IMM filter cannot realistically represent every possible target maneuver since doing so would require modeling a continuum of possibilities and adding new models increases computational cost. Moreover, adding models also distributes probability mass over more and more hypotheses, resulting in each model having less and less influence, eventually to a point where estimation accuracy can degrade. Another limitation is that the IMM assumes fixed Markov transition probabilities because tuning them online is typically too costly. Overall, the IMM is a good solution to the adaptive filtering problem, but its limitations can hinder estimation performance when tracking highly agile targets with complex dynamics.

Although there is some recent work [15–17] applying machine learning with recurrent or transformer architectures to either replace or augment traditional filters for tracking problems, none of these address the more challenging interception of maneuvering targets.

Our solution to improve filter performance for the scenario described in Section II is to augment a standard nine-state UKF with a target jerk estimate output from a policy optimized using reinforcement meta-learning (meta-RL). This jerk estimate is inserted into the filter's state propagation function as the rate of change for the target acceleration. Importantly, the policy has a recurrent network layer that allows mapping a *history* of observations to an estimate of target jerk. This enables the policy to roughly predict future changes in jerk direction and magnitude, which might be relatively small for a weaving target and much larger for a jinking target. Moreover, the recurrent policy can implement adaptive noise filtering. An advantage of predicting target jerk as opposed to selecting between multiple parameterizations of target maneuver types is that the latter encompasses a very large set of target maneuver possibilities, which can be more compactly modeled as a time series of target jerk predictions. Note that predicting jerk rather than acceleration allows the nine-state UKF to perform the work of tracking slowly and predictably varying components of target acceleration (such as drag force perpendicular to LOS) allowing the policy to focus on more challenging target acceleration components. As compared to other machine-learning based approaches, our approach has the advantage of leaving the standard UKF prediction and update equations in place, simplifying the flight certification process.

In previous work [18] we have found that to obtain performance targets in a high-fidelity simulator, it is necessary to optimize a state estimation system as part of the full guidance, navigation, and control (GN&C) system. Simplified simulators cannot capture the parasitic interactions between sub-systems, and the scale factor errors and non-Gaussian noise introduced through radome refraction and imperfect line of sight recovery. Consequently, in order to evaluate our approach, we have created a high-fidelity simulator to capture the most relevant noise sources and interactions between sub-systems. Where it is necessary (in order to avoid confusion) to explicitly indicate a variables reference frame, we use the superscripts "I", "B", "S", to denote the inertial, body, and LOS stabilized reference frames, respectively. To avoid overly complex notation, some sections do not use subscripts for all variables, but the reference frame is given at the start of the section. The paper is organized as follows. Section II describes the simulation environment and engagement scenarios, Section III develops the benchmark UKF and an IMM implementation. Section IV gives relevant background on RL and meta-RL, and formulates the flight control system (FCS) optimization problem in the meta-RL framework. Section V compares the performance of the meta-RL augmented UKF to that of the benchmark systems, and tests the ability of the jerk estimation policy to generalize to novel target maneuvers. Conclusions are given in Section VI.

## II. Simulation Environment

In this section, unless a variable has the "B" (body frame) or "S" (LOS stabilized frame) superscript, it is assumed to be in the inertial reference frame. Our simulator, shown in Fig. 1 uses an interceptor model that includes a guidance system, flight control system, an actuator model that models actuator bandwidth, fin limits, and fin rate limits, an inertial measurement unit (IMU) that models noise, bias, and placement distance from center of gravity, a radar receiver with elevation, azimuth, and range dependent noise, and explicit modeling of LOS recovery using a strap down seeker. The interceptor GN&C system must satisfy a load constraint, where the norm of the $y$ and $z$ body frame forces acting on the interceptor (see Section II.D) must be less than 75 g: $\|[F_y^{\mathrm{B}}, F_z^{\mathrm{B}}]\| < 75$ g, where g=9.81m/s$^2$ is the magnitude of the gravitational acceleration at the Earth's surface. The look angle constraint requires the maximum angle between the body frame x-axis and the body frame LOS vector to be less than 45°. Rather than directly imposing the look angle constraint by terminating an episode when the constraint is violated, we instead propagate the UKF forward for as long as the constraint is violated, without updating (as the measurements are invalid). Consequently, it is possible to violate the constraint for short periods and maintain good performance, noting that this constraint is always violated close to the end of the engagement.

### A. Engagement Scenario

Referring to Fig. 2*, the missile position vector, missile velocity vector, target position vector, and target velocity vector are shown as $\mathbf{r}_{\mathrm{M}}$, $\mathbf{v}_{\mathrm{M}}$, $\mathbf{r}_{\mathrm{T}}$, $\mathbf{v}_{\mathrm{T}}$, respectively. We can also define the relative position and velocity vectors $\mathbf{r}_{\mathrm{TM}} = \mathbf{r}_{\mathrm{T}} - \mathbf{r}_{\mathrm{M}}$ and $\mathbf{v}_{\mathrm{TM}} = \mathbf{v}_{\mathrm{T}} - \mathbf{v}_{\mathrm{M}}$. The elevation angle $\theta_E$ is the angle between $\mathbf{r}_{\mathrm{TM}}$ and its projection onto the x-y plane. We randomly generate the target's initial velocity vector such that $\mathbf{v}_{\mathrm{T}}$ lies within a cone with axis $\mathbf{r}_{\mathrm{TM}}$ and half apex angle $\theta_{\mathbf{v}_{\mathrm{T}}}$. In $\mathbb{R}^3$ we define a heading error ($HE$) as the angle between the interceptor's initial velocity vector and the velocity vector associated with the lead angle required to put the interceptor on a collision heading with the target. Note that due to the interceptor aerodynamic forces and target acceleration, this is not a perfect collision triangle, and the true heading error is greater than $HE$.

During the midcourse guidance phase, ground radar tracks both the interceptor and target and periodically transmits track updates of the relative state to the interceptor, before handing off target tracking to the interceptor's active radar

*In this figure, the illustrated vectors are not in general within the x-z plane

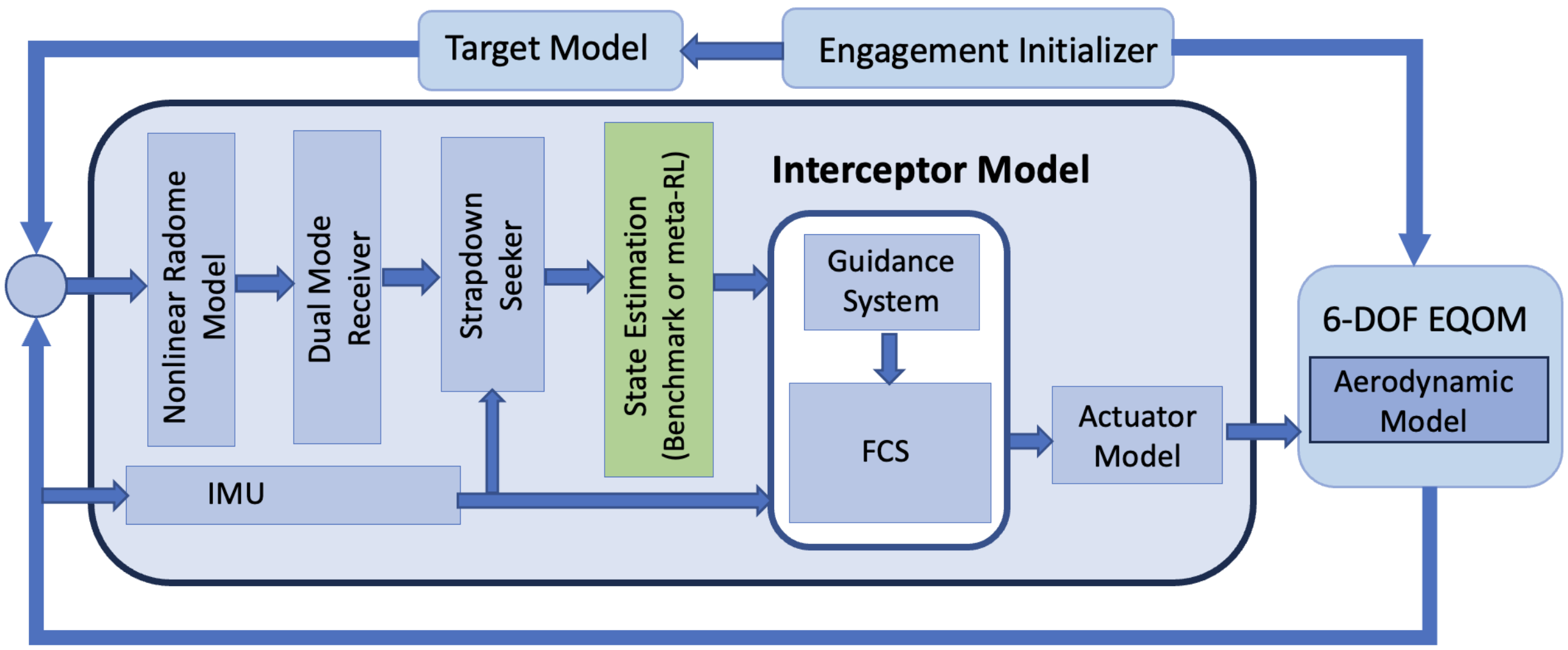

**Fig. 1 Simulator Diagram**

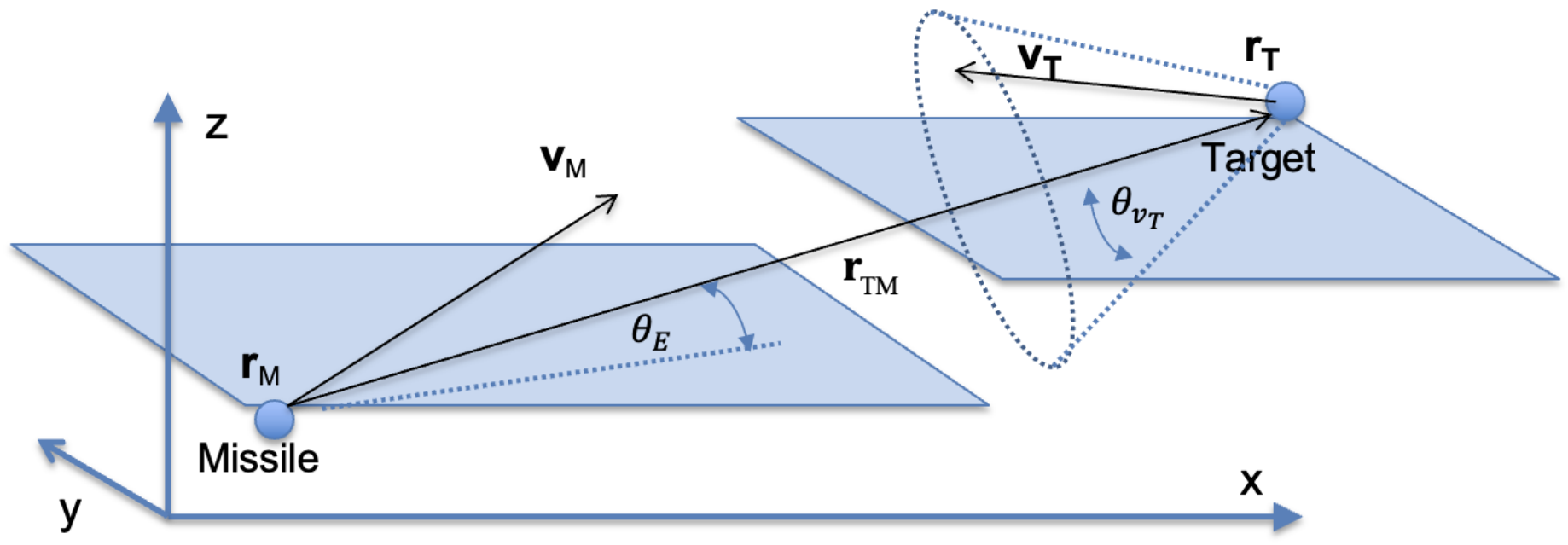

**Fig. 2 Engagement**

seeker during the terminal phase. Consequently, the state estimator used for the terminal phase of the engagement only needs to work over the "basket" of initial conditions consistent with the end of the mid-course guidance phase. This basket is limited by a minimum range-to-target that gives the interceptor enough time to complete a successful terminal phase intercept, with the minimum time being a function of target maneuver, closing speed, and initial heading error. However, due to unpredictable target behavior between track updates, there will be some amount of uncertainty regarding the interceptor's state at the start of the terminal phase. The bounds on the initial conditions and initial filter uncertainty are given in Table 1. In Section V.C we expand the initial condition uncertainty and filter initial uncertainty to test filter generalization capability.

### B. Target Evasive Acceleration Model

Both the jerk estimation policy and IMM benchmark were optimized using two types of evasive maneuvers, a weaving maneuver and a jinking maneuver. The weaving maneuver has randomized phase offset and frequency, with frequency shared for each axis and phase independent for each axis in $\mathbb{R}^3$, whereas with the jinking maneuver, a given axis may or may not have a transition at the jink frequency, but the transitions on each axis are in phase. Each of these maneuvers has a 1/2 probability of being selected at the start of each episode. At each simulation step, the target's evasive acceleration vector is modified by removing the component parallel with the target velocity vector and then adjusted for dynamic pressure, with a nominal acceleration capability of 4 g at an altitude of 20 km and a speed of 2500 m/s. Thus, the target acceleration vector is always orthogonal to the target velocity. The maneuvers have varying acceleration levels and random start time, duration, and switching time. At the start of each episode, with probability

**Table 1 Simulation Initial Conditions)**

| Parameters Drawn Uniformly | Min | Max |
|---|---|---|
| State Estimator Position Initial Uncertainty (%) | -10 | 10 |
| State Estimator Velocity Initial Uncertainty (%) | -10 | 10 |
| State Estimator Acceleration Initial Uncertainty (%) | -30 | 30 |
| Elevation Angle to Target $\theta_E$ (degrees) | 30 | 60 |
| Azimuth Angle to Target $\theta_A$ (degrees) | -180 | 180 |
| Altitude (m) | 15000 | 22000 |
| Missile Velocity Magnitude $\|\mathbf{v}_{\mathrm{M}}\|$ (m/s) | 2300 | 2600 |
| Heading Error (degrees) | 0 | 3 |
| Initial Angle of Attack $\alpha$ (degrees) | -5 | -5 |
| Initial Side Slip Angle $\beta$ (degrees) | -5 | 5 |
| Initial Range to Targeted Asset $\|\mathbf{r}_{\mathrm{AT}}\|$ (m) | 20000 | 25000 |
| Target Velocity Magnitude $\|\mathbf{v}_{\mathrm{T}}\|$ (m/s) | 3000 | 4000 |
| Target Velocity Cone Half Apex Angle $\theta_{\mathbf{v}_{\mathrm{T}}}$ (degrees) | 0 | 15 |
| Interceptor Maximum Acceleration (@ 2500 m/s 20000 m, applied to each component) (g) | - | 20 |
| Target Evasion Acceleration (@ 2500 m/s 20000 m, applied to each component) (g) | 0 | 4 |
| Target Weave Period (s) | 0.5 | 4.0 |

0.5 the maneuvers in that episode use the full evasion acceleration capability (4 g at 2500 m/s and 20 km), and with probability 0.5 the acceleration is scaled by a factor uniformly distributed between 0 and 1. Note that target drag orthogonal to the interceptor LOS is also perceived as a target maneuver; the target drag model is given in Section II.E. Sample target maneuvers are shown in Fig. 3, which show evasive acceleration only (not drag orthogonal to LOS), and include a step maneuver that is used as one of the maneuvers to test filter generalization performance in Section V.C.

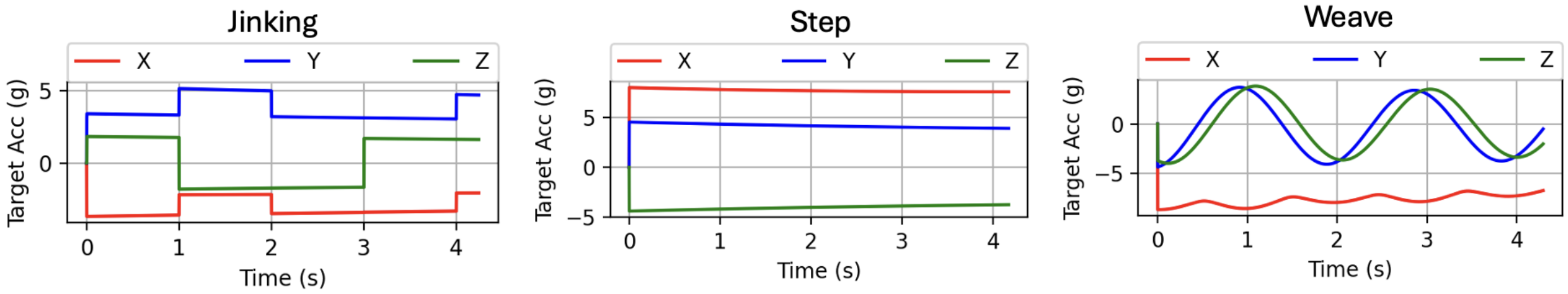

**Fig. 3 Sample Target Maneuvers (only evasive acceleration shown**

## C. Interceptor Components

*1. Radome Model*

We calculate the refracted azimuth $\tilde{\theta}_u$ and elevation $\tilde{\theta}_v$ angles as shown in Eqs. (1a) through (1b), where our nominal value of $A$ is 0.5 mrad, and $K_u$, $K_v$ are set to 3.0. Note we assume that there is a combination of software (such as compensation lookup tables) and hardware (such as radome shaping) that reduce the effective radome refraction to these values.

$$\tilde{\theta}_u = \theta_u + A\left(0.75\frac{\theta_u}{\pi/2} + 0.25\cos\left(\frac{2\pi}{k_u}\theta_u\right)\right) \tag{1a}$$

$$\tilde{\theta}_v = \theta_v + A\left(0.75\frac{\theta_v}{\pi/2} + 0.25\cos\left(\frac{2\pi}{k_v}\theta_v\right)\right) \tag{1b}$$

The refraction angle $\theta_R$ is shown in Fig. 4 for various values of $k$. Note that the radome slope $\frac{\partial \theta_R}{\partial \theta_L}$ is given by the slope of the curves in the figure.

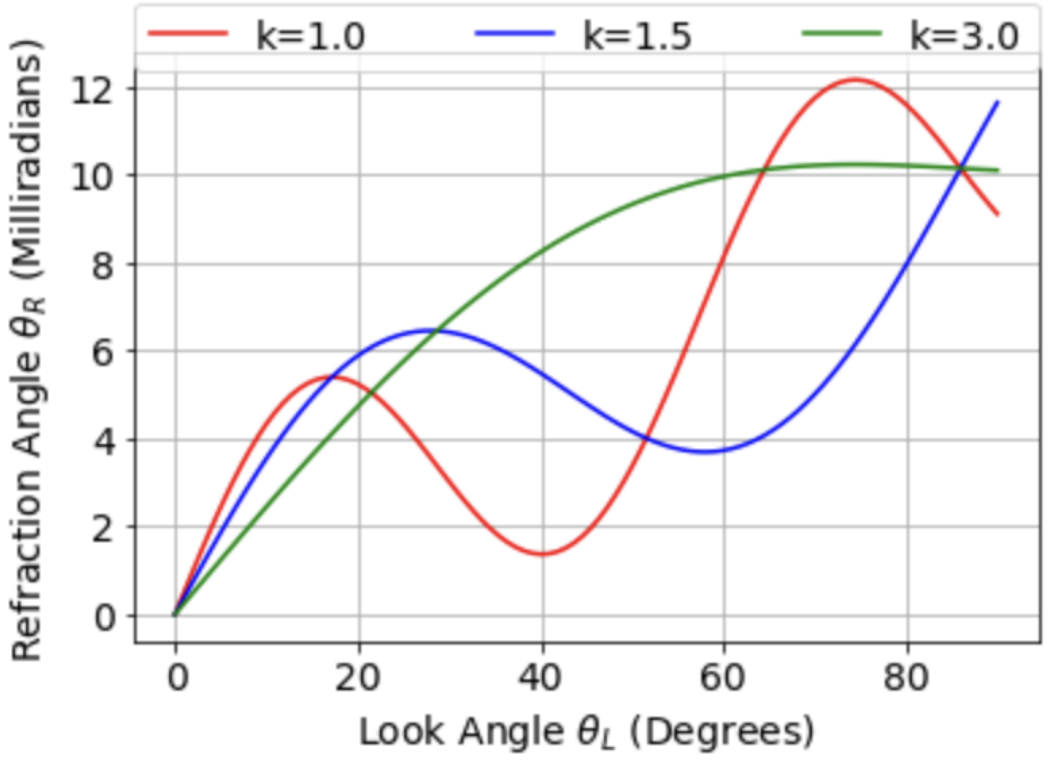

**Fig. 4 Radome Refraction Angle as Function of Look Angle**

*2. Seeker Model*

We model sensor noise (elevation, azimuth, and range) as Gaussian noise, using a standard deviation of 1% of range for range dependent range noise and 0.2 mrad for angle noise. Obtaining this level of angle noise requires a dual model seeker (radar for range, IR for angles), similar to the Israeli Stunner interceptor. This improves angular resolution, target discrimination, and robustness to countermeasures. We do not separately model the IR seeker window and radome.

*3. Rate Gyro Model*

Our rate gyro model corrupts the ground truth rotational velocity $\boldsymbol{\omega}^{\text{B}}$ with both scale factor bias and Gaussian noise, as shown in Eq. (2a), where $\mathcal{U}(a, b, n)$ denotes an $n$ dimensional uniformly distributed random variable bounded by $(a, b)$ that is generated at the start of each episode, and each dimension of the random variable is independent. We use $\boldsymbol{\epsilon}_\omega = 0.001$ and $\sigma_\omega = 0.1°$. The rotational velocity is passed through a two pole low pass filter with a 100Hz natural frequency and 0.707 damping, as shown in Eq. 2b.

$$\boldsymbol{\omega}_1^{\text{B}} = \boldsymbol{\omega}(1 + \mathcal{U}(-\boldsymbol{\epsilon}_\omega, \boldsymbol{\epsilon}_\omega, 3)) \tag{2a}$$

$$\boldsymbol{\omega}_{\text{obs}}^{\text{B}} = \text{LPF}(\boldsymbol{\omega}_{\mathbf{1}}^{\text{B}}) + \mathcal{N}(0, \sigma_\omega, 3) \tag{2b}$$

*4. Accelerometer Model*

We account for accelerometer placement using Eq. (3a), with $\mathbf{c} = \mathbf{r}_{\text{cg}} - \mathbf{r}_{\text{acc}}$, where $\mathbf{r}_{\text{acc}}$ is the accelerometer location which we assume to be 25cm back from the nose. Similar to the rate gyro model, the body frame acceleration is corrupted with a scale factor error as shown in Eqs. (3b) and (3c), where we use $\boldsymbol{\epsilon}_{\text{acc}} = 0.001$, and "LPF" models the accelerometer dynamics as a two pole low pass filter with a 100Hz natural frequency and 0.707 damping, and $\sigma_{\text{acc}} = 0.01$ m/s$^2$.

$$\mathbf{a}_1^{\text{B}} = \mathbf{a}^{\text{B}} + [\dot{\omega}\times]\mathbf{c} + [\omega\times][\omega\times](\mathbf{c} + \delta\mathbf{c}) \tag{3a}$$

$$\mathbf{a}_2^{\text{B}} = \text{LPF}(\mathbf{a}_1^{\text{B}}) \tag{3b}$$

$$\mathbf{a}_{\text{obs}}^{\text{B}} = \mathbf{a}_2^{\text{B}}(1 + \mathcal{U}(-\boldsymbol{\epsilon}_{\text{acc}}, \boldsymbol{\epsilon}_{\text{acc}}, 3)) + \mathcal{N}(0, \sigma_{\text{acc}}, 3) \tag{3c}$$

*5. Line of Sight Recovery*

Referring to Fig. 1, the refracted body frame line of sight unit vector $\hat{\mathbf{r}}^{\mathrm{B}}_{\mathrm{TM}}$ from the dual-mode receiver block is input to the LOS recovery block. This block uses the estimated change of attitude since the start of the engagement $\mathbf{dq}_{\mathrm{obs}}$ to rotate $\hat{\mathbf{r}}^{\mathrm{B}}_{\mathrm{TM}}$ into the reference frame corresponding to the interceptor's attitude at the start of the engagement. Specifically, at each guidance step we map $\mathbf{dq}_{\mathrm{obs}}$ to a direction cosine matrix $\mathbf{C}(\mathbf{dq}_{\mathrm{obs}})$, and compute the recovered (stabilized) LOS as $\hat{\mathbf{r}}^{\mathrm{S}}_{\mathrm{TM}} = [\mathbf{C}(\mathbf{dq}_{\mathrm{obs}})]^{\mathrm{T}}\hat{\mathbf{r}}^{\mathrm{B}}_{\mathrm{TM}}$.

The estimated change in attitude $\mathbf{dq}_{\mathrm{obs}}$ is parameterized as a quaternion, and is computed by integrating $\boldsymbol{\omega}$ as shown in Equation 4, where $\mathbf{dq}_{\mathrm{obs}}$ is reset at the start of each episode to $[1, 0, 0, 0]$. In our simulation model, we approximate this integration using fourth-order Runge-Kutta integration [19]. In the unrealistic case where we have access to the ground truth value of $\boldsymbol{\omega}$, this completely decouples $\hat{\mathbf{r}}_{\mathrm{TM}}$ from body rotation. However, since $\boldsymbol{\omega}_{\mathrm{obs}}$ is corrupted with a scale factor error and Gaussian noise, in general $\mathbf{dq}_{\mathrm{obs}} \neq \mathbf{dq}$, and the LOS recovery will be imperfect. The combination of imperfect LOS recovery and radome refraction result in a false indication of target motion. This results in increased miss distance, and can potentially destabilize the G&C system. This is discussed in more detail in [20–22].

$$\begin{bmatrix} \dot{\mathrm{dq}}_0 \\ \dot{\mathrm{dq}}_1 \\ \dot{\mathrm{dq}}_2 \\ \dot{\mathrm{dq}}_3 \end{bmatrix} = \frac{1}{2} \begin{bmatrix} \mathrm{dq}_0 & -\mathrm{dq}_1 & -\mathrm{dq}_2 & -\mathrm{dq}_3 \\ \mathrm{dq}_1 & \mathrm{dq}_0 & -\mathrm{dq}_3 & \mathrm{dq}_2 \\ \mathrm{dq}_2 & \mathrm{dq}_3 & \mathrm{dq}_0 & -\mathrm{dq}_1 \\ \mathrm{dq3} & -\mathrm{dq}_2 & \mathrm{dq}_1 & \mathrm{dq}_0 \end{bmatrix} \begin{bmatrix} 0 \\ \omega_{\mathrm{obs}_0} \\ \omega_{\mathrm{obs}_1} \\ \omega_{\mathrm{obs}_2} \end{bmatrix} \tag{4}$$

*6. Guidance System*

The guidance system implements true proportional navigation (TPN) [23] to calculate a commanded acceleration $\mathbf{a}^{\mathrm{S}}$ from the LOS rotation rate, as shown in Eqs. (5a) through (5c). This commanded acceleration is then rotated into the body frame using the estimated change in attitude $\mathbf{dq}$ and adjusting using g-limits. These g-limits assume a maximum acceleration of 25 g at an altitude of 20 km and a speed of 2500 m/s, and are adjusted for dynamic pressure. This body frame commanded commanded acceleration is then used to compute the acceleration tracking error according to $\mathbf{a}^{\mathrm{B}}_{\mathrm{err}} = \mathbf{a}_{\mathrm{com}}[y, z]^{\mathrm{B}} - \mathbf{a}_{\mathrm{obs}}[y, z]^{\mathrm{B}}$, where $\mathbf{a}^{\mathrm{B}}_{\mathrm{obs}}$ is the accelerometer measurement and $[y, z]$ indicates we are computing the error over the body frame $y$ and $z$ components. We use $N = 4$ for the navigation ratio. Note that the "S" and "B" superscripts indicate the LOS stabilized (see Section II.C.5) and body frame reference frames, respectively.

$$\boldsymbol{\Omega} = \frac{\mathbf{r}^{\mathrm{S}}_{\mathrm{TM}} \times \mathbf{v}^{\mathrm{S}}_{\mathrm{TM}}}{\mathbf{r}^{\mathrm{S}}_{\mathrm{TM}} \cdot \mathbf{r}^{\mathrm{S}}_{\mathrm{TM}}} \tag{5a}$$

$$\mathbf{a}^{\mathrm{S}}_{\mathrm{com}} = -N v_{c_{\mathrm{obs}}} \left( \frac{\mathbf{r}^{\mathrm{S}}_{\mathrm{TM}}}{\|\mathbf{r}^{\mathrm{S}}_{\mathrm{TM}}\|} \times \boldsymbol{\Omega} \right) \tag{5b}$$

$$\mathbf{a}^{\mathrm{B}}_{\mathrm{com}} = [\mathbf{C}(\mathbf{dq}_{\mathrm{obs}})]\mathbf{a}^{\mathrm{S}}_{\mathrm{com}} \tag{5c}$$

*7. Flight Condition Estimation*

The FCS requires an estimate of angle of attack $\alpha$, sideslip angle $\beta$, altitude $h$, and Mach $M$. These observations are generated by integrating $\mathbf{a}^{\mathrm{B}}_{\mathrm{obs}}$ to obtain an estimate of body frame velocity and position, allowing computation of estimated interceptor velocity $\mathbf{v}^{\mathrm{B}}_{\mathrm{obs}}$ and position $\mathbf{r}^{\mathrm{B}}_{\mathrm{obs}}$, with observed angle of attack $\alpha_{\mathrm{obs}}$ and sideslip angle $\beta_{\mathrm{obs}}$ computed from $\mathbf{v}^{\mathrm{B}}_{\mathrm{obs}}$. $\mathbf{r}^{\mathrm{B}}_{\mathrm{obs}}$ is rotated into the inertial frame using estimated attitude, allowing computation of altitude $h_{\mathrm{obs}}$. Mach $M$ is computed from estimated altitude and speed using an interpolated Earth atmospheric model table. Finally, for roll stabilization, the FCS requires an estimate of roll ($\theta_{\mathrm{roll}_{\mathrm{obs}}}$), which is computed by integrating $\boldsymbol{\omega}_{\mathrm{obs}_x}$.

*8. Flight Control*

In order to give the best comparison between the benchmarks and the RL-ABF, we use the flight control system (FCS) optimized using meta-RL described in [24], where, we show that the meta-RL optimized FCS achieves a significantly higher hit probability as compared to a 3-loop autopilot with tuned gain tables, allowing a better comparison between state estimation methods. The flight control system maps $\begin{bmatrix} \mathbf{a}^{\mathrm{B}}_{\mathrm{err}} & \boldsymbol{\omega}_{\mathrm{obs}} & \alpha_{\mathrm{obs}} & \beta_{\mathrm{obs}} & M & h_{\mathrm{obs}} & \theta_{\mathrm{roll}_{\mathrm{obs}}} \end{bmatrix}$ to the commanded channel deflections (see Section II.C.9 ).

*9. Actuator Model*

The channel commanded deflections $\boldsymbol{\delta}_\mathrm{C} = [\delta_\mathrm{P}, \delta_\mathrm{Y}, \delta_\mathrm{A}, \delta_\mathrm{SB}]$ (pitch, yaw, roll, and speed brake channel commands) from the FCS are converted to physical fin deflections $\boldsymbol{\delta}_\mathrm{F} = [\delta_1, \delta_2, \delta_3, \delta_4]$ using the fin mix matrix shown in Eq. (6), and the transformations shown in Eqs. (7a) and (7b). Note that the speed brake is unused.

$$\mathbf{M}_\mathrm{fin} = \begin{bmatrix} 1 & 0 & 1 & -1 \\ 0 & 1 & 1 & 1 \\ -1 & 0 & 1 & -1 \\ 0 & -1 & 1 & 1 \end{bmatrix} \tag{6}$$

$$\boldsymbol{\delta}_\mathrm{C} = \mathbf{M}_\mathrm{fin}^{-1}\boldsymbol{\delta}_\mathrm{F} \tag{7a}$$

$$\boldsymbol{\delta}_\mathrm{F} = \mathbf{M}_\mathrm{fin}\boldsymbol{\delta}_\mathrm{C} \tag{7b}$$

## D. Aerodynamic Model

The interceptor geometry is taken from [25] and is shown in Fig. 5, with the nominal center of gravity set to -0.54 m. Our nominal aerodynamic model is derived from slender body and Newtonian impact theory [26, 27]. The method in [27] is extended to 6-DOF using a method similar to that in [28], but with a higher fidelity axial force model, rolling moment model, and the addition of a cross channel interference term for control effectiveness. The aerodynamics are similar to that obtained using Missile DATCOM in other work [18]. The vehicle diameter is $d = 0.105$ m. The length is $x_\mathrm{L} = 1.05$ m, and the inertia tensor is given in Figure 8. The vehicle mass is 13.5 kg, the reference length is $\mathrm{L}_\mathrm{ref} = d$, and the reference area is $\mathrm{S}_\mathrm{ref} = \pi(d/2)^2$. A sample open loop response to flight control commanded deflections are shown in Fig. 6, where $\boldsymbol{\delta}_\mathrm{C} = \delta_\mathrm{P}, \delta_\mathrm{Y}, \delta_\mathrm{A}$ are the FCS commanded channel deflections as described in Section II.C.9. We see that the vehicle is slightly open-loop unstable.

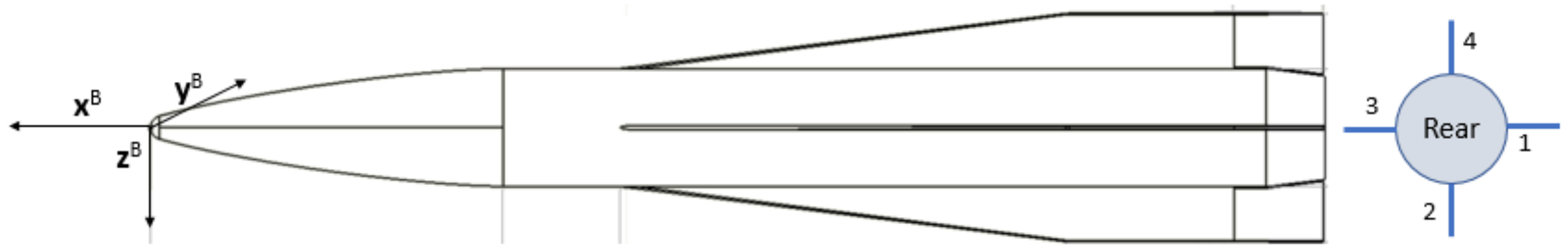

**Fig. 5 Interceptor Geometry**

$$\mathbf{J} = \begin{bmatrix} 0.02 & 0 & 0 \\ 0 & 0.95 & 0 \\ 0 & 0 & 0.95 \end{bmatrix} \tag{8}$$

We model aerodynamic uncertainty by randomly modifying the shape of the aerodynamic coefficient curves for each simulated trajectory. To do so, we first bias the coefficients returned from the nominal aerodynamic model as follows. At the start of each episode with probability $p = 0.5$ we uniformly sample an aerodynamic coefficient bias $\boldsymbol{\epsilon}_\mathrm{coeff} \in \mathbb{R}^4$ between some bounds $+/-\epsilon$, otherwise a coefficient bias is (independently) set to either $-\epsilon$ or $\epsilon$ with equal probability, noting that in both cases the bias applied to the coefficient groups ($C_z$,$C_y$), ($C_m$,$C_n$), $C_x$, and $C_l$ are independent. We assume that the bias on $C_z$ and $C_y$ is identical to symmetry (and similarly for $C_m$ and $C_n$). Then, for the remainder of the episode, the force and moment coefficients are biased using the sampled value of $\boldsymbol{\epsilon}_\mathrm{coeff}$. We account for uncertainty in the center of gravity (CG) and center of pressure (CP) by modifying the moments as shown in Eqs. (9a) through (9d). Here $\boldsymbol{\Delta}_\mathrm{CP}$ is uniformly generated at the start of each episode and MRP is the moment reference point, which in this work is set to the CG. Due to symmetry, errors in $\Delta_{\mathrm{CP}_y}$ and $\Delta_{\mathrm{CP}_z}$ will mostly cancel. We use $\boldsymbol{\epsilon}_\mathrm{coeff} = 0.10$ and $\Delta_{\mathrm{CGCP}_{x,y,z}} = 0.01$ m in all experiments.

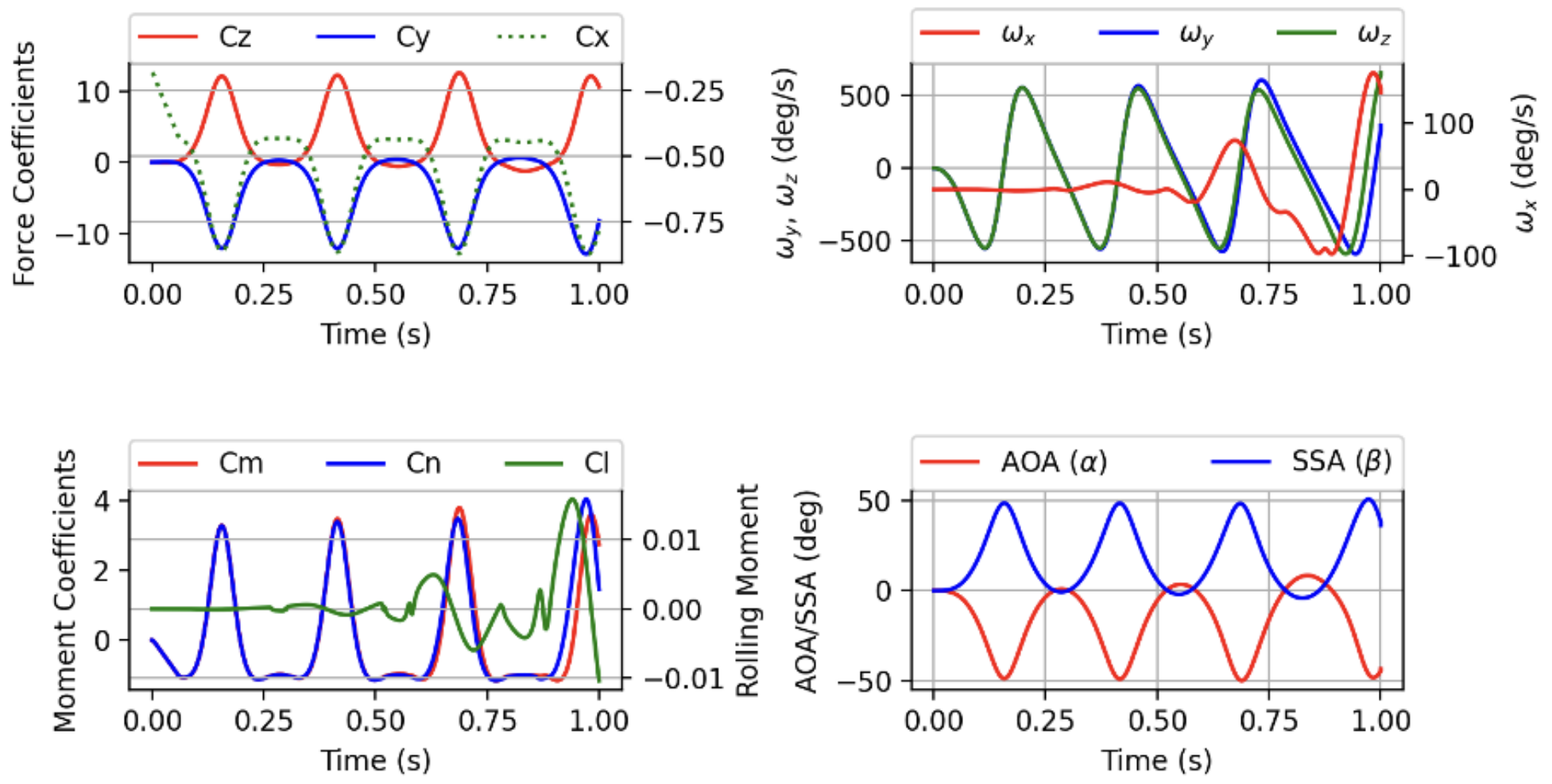

**Fig. 6 Open Loop Airframe Response $\delta_C = [30, 30, 0]$, Altitude=25 km, speed=2500 m/s**

$$\Delta_{\text{CGCP}} = \frac{(\text{MRP} - \text{CG}) + \Delta_{\text{CP}}}{L_{\text{ref}}} \times [Cx, Cy, Cz] \tag{9a}$$

$$Cl_{\text{adj}} = Cl + \Delta_{\text{CGCP}_x} \tag{9b}$$

$$Cm_{\text{adj}} = Cm + \Delta_{\text{CGCP}_y} \tag{9c}$$

$$Cn_{\text{adj}} = Cn + \Delta_{\text{CGCP}_z} \tag{9d}$$

### E. Equations of Motion

Let $\mathbf{v}^{\text{B}} \equiv [u, v, w]$ denote the body frame interceptor velocity. The angle of attack $\alpha$ and sideslip angle $\beta$ are then computed as shown in Eqs. (10a) through (10b).

$$\alpha = \arctan \frac{w}{u} \tag{10a}$$

$$\beta = \arctan \frac{v}{u} \tag{10b}$$

(10c)

The atmospheric density $\rho$ and Mach number $M$ are calculated using the 1976 standard Earth atmospheric temperature model. The interceptor control surface deflections $\boldsymbol{\theta}_{\text{ctrl}} = [\theta_{\text{LT}}, \theta_{\text{RT}}, \theta_{\text{BT}}, \theta_{\text{UT}}]$, dynamic pressure $q_o = 0.5\rho V^2$ with $V = \|\mathbf{v}^B\|$, rotational velocity vector $\omega$, $M$, $\alpha$ and $\beta$ are then input to the aerodynamic model described in Section II.D to compute the body frame force $\mathbf{F}^{\text{B}}$ and torque $\mathbf{L}^{\text{B}}$.

The rotational velocity $\boldsymbol{\omega}$ is updated by integrating the Euler rotational equations of motion, as shown in Equation (11), where $\mathbf{J}$ is the interceptor's inertia tensor, and with the skew symmetric operator $[\mathbf{a}\times]$ defined in Eq. (12).

$$\mathbf{J}\dot{\omega} = -[\omega\times]\mathbf{J}\omega + \mathbf{L}^{\text{B}} \tag{11}$$

$$[\mathbf{a}\times] \equiv \begin{bmatrix} 0 & -a_3 & a_2 \\ a_3 & 0 & -a_1 \\ -a_2 & a_1 & 0 \end{bmatrix} \tag{12}$$

The interceptor's attitude $\mathbf{q}$ is updated by integrating the differential kinematic equations shown in Equation (13), where the interceptor's attitude is parameterized using the quaternion representation and $\boldsymbol{\omega}_i$ denotes the $i^{th}$ component of the rotational velocity vector $\boldsymbol{\omega}_B$.

$$\begin{bmatrix}\dot{q}_0\\ \dot{q}_1\\ \dot{q}_2\\ \dot{q}_3\end{bmatrix} = \frac{1}{2}\begin{bmatrix} q_0 & -q_1 & -q_2 & -q_3\\ q_1 & q_0 & -q_3 & q_2\\ q_2 & q_3 & q_0 & -q_1\\ q_3 & -q_2 & q_1 & q_0\end{bmatrix}\begin{bmatrix}0\\ \omega_0\\ \omega_1\\ \omega_2\end{bmatrix} \tag{13}$$

The interceptor body frame velocity $\mathbf{v}^{\mathrm{B}}$ is updated by integrating the differential equation given in Eq. 14, where $m$ is the interceptor mass, $\mathbf{C}_{\mathrm{BN}}(\mathbf{q})$ is the DCM mapping from the inertial to body frame given the interceptor attitude $\mathbf{q}$, and $g = -9.81$ m/s$^2$.

$$\dot{\mathbf{v}}^{\mathrm{B}} = -[\omega\times]\mathbf{v}^{\mathrm{B}} + \frac{\mathbf{F}^{\mathrm{B}}}{m} + [\mathbf{C}_{\mathrm{BN}}(\mathbf{q})]^{\top}\begin{bmatrix}0 & 0 & g\end{bmatrix} \tag{14}$$

$\mathbf{v}^{\mathrm{B}}$ is then rotated into the North-East-Down (NED) inertial frame as shown in Eq. (15a), and $\mathbf{v}_{\mathrm{M}}$ is calculated by rotating the vector into the East-North-Up (ENU) reference frame depicted in Fig. 2.

$$\mathbf{v}^{\mathrm{N}} = \mathbf{C}_{\mathrm{BN}}(\mathbf{q})\mathbf{v}^{\mathrm{B}} \tag{15a}$$

$$\mathbf{v}_{\mathrm{M}} = \mathbf{C}_{\mathrm{NE}}\mathbf{v}^{\mathrm{N}} \tag{15b}$$

The interceptor inertial frame position $\mathbf{r}_{\mathrm{M}}$ is then updated by integrating Eq. (16)

$$\dot{\mathbf{r}}_{\mathrm{M}} = \mathbf{v}_{\mathrm{M}} \tag{16}$$

The target inertial frame position $\mathbf{r}_{\mathrm{T}}$ and velocity unadjusted for drag $\tilde{\mathbf{v}}_{\mathrm{T}}$ are then updated by integrating Eqs. (17a) and (17b). The target speed $V_{\mathrm{T}} = \|\mathbf{v}_{\mathrm{T}}\|$ is adjusted for drag by integrating Eq. (17c), where at the start of each episode we randomly generate $k_{\mathrm{T}}$ from 0.1 to 0.25 and $\mathrm{cd}_0$ from 0.05 to 0.10. We use $m_{\mathrm{T}} = 450$ kg.

$$\dot{\mathbf{r}}_{\mathrm{T}} = \mathbf{v}_{\mathrm{T}} \tag{17a}$$

$$\dot{\tilde{\mathbf{v}}}_{\mathrm{T}} = \mathbf{a}_{\mathrm{T}} \tag{17b}$$

$$\dot{V}_{\mathrm{T}} = -\frac{\rho(h_{\mathrm{T}})V_{\mathrm{T}}^2\mathrm{cd}_0}{2m_{\mathrm{T}}} - k_{\mathrm{T}}\|\mathbf{a}_{\mathrm{T}}\| \tag{17c}$$

The target velocity is then computed as shown in Eq. (18)

$$\mathbf{v}_{\mathrm{T}} = V_{\mathrm{T}}\frac{\tilde{\mathbf{v}}_{\mathrm{T}}}{\|\tilde{\mathbf{v}}_{\mathrm{T}}\|} \tag{18}$$

The equations of motion are updated using fourth-order Runge-Kutta integration. For ranges greater than 50 m, a timestep of 5 ms is used, and for the final 50 m of homing, a timestep of 25 us is used in order to more accurately (within 0.125 m) measure miss distance; this technique is borrowed from [29].

# III. Benchmarks

In this section, all variables are assumed to be in the LOS stabilized reference frame (see Section II.C.5 unless otherwise indicated with the "B" subscript for body frame.

## A. Unscented Kalman Filter Benchmark

We use the UKF [5, 6] to create benchmarks for performing relative state estimation. Our UKF benchmarks use the dynamics and measurement models shown in Eqs. (19a) and (19b)

$$\mathbf{x}_k = \boldsymbol{f}_k(\mathbf{x}_{k-1}) + \boldsymbol{\upsilon}_k \tag{19a}$$

$$\mathbf{y}_k = \boldsymbol{h}_k(\mathbf{x}_{k|k-1}) + \boldsymbol{\eta}_k \tag{19b}$$

Here $\mathbf{x}_k \in \mathbb{R}^{n_x}$ is the state, $\boldsymbol{f}_k : \mathbb{R}^{n_x} \rightarrow \mathbb{R}^{n_x}$ is a known discrete dynamics propagation function that propagates from time step $k-1$ to time step $k$ using fourth-order Runge-Kutta integration, and $\boldsymbol{\upsilon}_k \in \mathbb{R}^{n_x}$ is zero mean process noise with covariance $\mathbf{Q}_k$. Additionally, $\mathbf{y} \in \mathbb{R}^{n_y}$ is the measurement, $\boldsymbol{h}_k : \mathbb{R}^{n_x} \to \mathbb{R}^{n_y}$ is a known measurement function, and $\boldsymbol{\eta}_k \in \mathbb{R}^{n_y}$ is zero mean measurement noise with covariance $\mathbf{R}_k$, uncorrelated with all other noise sources. $\mathbf{Q}$ characterizes uncertainty in the system dynamics and increases the predicted state covariance to account for modeling errors and unmodeled disturbances, while $\mathbf{R}$ characterizes uncertainty in the sensor measurements and determines how strongly the filter weights new measurements relative to the predicted state during the update.

The UKF benchmark uses nine states such that $\mathbf{x} = [\mathbf{r}_{\text{TM}},\ \mathbf{v}_{\text{TM}},\ \mathbf{a}_{\text{T}}]$, where $\mathbf{r}_{\text{TM}} \in \mathbb{R}^3$ is the relative position, $\mathbf{v}_{\text{TM}} \in \mathbb{R}^3$ is the relative velocity , and $\mathbf{a}_{\text{T}} \in \mathbb{R}^3$ is the target acceleration. The state propagation function is shown in Eqs. (21a) through 21c.

$$\dot{\mathbf{r}}_{\text{TM}} = \mathbf{v}_{\text{TM}} \tag{20a}$$
$$\dot{\mathbf{v}}_{\text{TM}} = \mathbf{a}_{\text{T}} - \mathbf{a}_{\text{M}} \tag{20b}$$
$$\dot{\mathbf{a}}_{\text{T}} = \mathbf{0} \tag{20c}$$

Here $\mathbf{a}_{\text{M}} = [\mathbf{C}(\mathbf{dq}_{\text{obs}})]^\top \mathbf{a}^{\text{B}}_{\text{obs}}$, and $\mathbf{a}^{\text{B}}_{\text{obs}}$ is the interceptor's acceleration as sensed by the IMU. When we investigate meta-RL for jerk estimation, the zero vector in Eq. (20c) is replaced with the estimate for target jerk.

The measurement function $\boldsymbol{h}_k$ is shown in Eqs. (21a)–(21c), which computes the predicted azimuth, elevation, and range using zero-based indexing.

$$\phi = \text{atan2}([\mathbf{r}_{\text{TM}}]_1, [\mathbf{r}_{\text{TM}}]_0) \tag{21a}$$
$$\theta = \text{atan2}([\mathbf{r}_{\text{TM}}]_2, r) \tag{21b}$$
$$r = \|\mathbf{r}_{\text{TM}}\| \tag{21c}$$

The measurement covariance matrix $\mathbf{R}_k$ is shown in Eq. (22), where $\theta_{\text{std}}$ and $\phi_{\text{std}}$ are the standard deviations (std) of the elevation and azimuth measurements, whereas $r_{\text{std}}$ is multiplied by the current range measurement $r$ to give a range-dependent standard deviation:

$$\mathbf{R}_k = \begin{bmatrix} \theta^2_{\text{std}} & 0 & 0 \\ 0 & \phi^2_{\text{std}} & 0 \\ 0 & 0 & r^2_{\text{std}}\, r \end{bmatrix} \tag{22}$$

We use $\theta_{\text{std}} = 0.002$, $\phi_{\text{std}} = 0.002$, and $r_{\text{std}} = 0.2$. Recall that the actual standard deviation of the range-dependent noise in the seeker model (see Section II.C.2) is 1% of range, which implies a lower covariance than $0.2^2$, whereas the remainder is from imperfect LOS recovery and radome refraction. Note, the initial values of $\theta^2_{\text{std}}$, $\phi^2_{\text{std}}$, and $r^2_{\text{std}}\, r$ are found by viewing UKF estimation plots from sample trajectories, and then tuned via a grid search over Monte Carlo simulations.

We implement the UKF benchmarks using the `filterpy` Python package. We experimented with different UKF process noise models, using near constant acceleration (nca) [1], near constant jerk (ncj) [2], and Singer (nas) [3]. Each of the process noise matrices is parameterized with a standard deviation $Q_{\text{std}}$. We ran a rough grid search to find the best process noise standard deviation $q_{\text{std}}$. These filters are compared in Table 2, where the "Miss < 25cm" field gives the percentage of episodes that resulted in miss distance less than 25cm, which we consider to be a direct impact on the target. The "RMS $\|\mathbf{a}_{\text{M}}\|$" field gives the body frame x,y components of $\|\mathbf{a}_{\text{M}}\|$, a control effort metric, whereas the "RMS $\|\mathbf{a}^{\text{E}}_{\text{T}}\|$" field shows the target acceleration prediction error. Finally the "Vio" field gives the percentage of episodes that resulted in a load constraint violation. Although performance is close for all the nine-state filters, the "nas" process noise model has the best overall performance. Note that we also experimented with a twelve-state UKF, but found that this significantly underperformed the nine-state filters. This is likely due to the poor observability of target jerk using only angle and range measurements.

**Table 2 Target Maneuver Comparison**

| Process Noise | Miss < 25cm (%) | RMS $\|\mathbf{a}_{\text{M}}\|$ (m/s$^2$) | RMS $\|\mathbf{a}^{\text{E}}_{\text{T}}\|$ (m/s$^2$) | Vio (%) |
|---|---|---|---|---|
| nca ($q_{\text{std}} = 3$) | 39 | 47 | 82 | 0.0 |
| ncj ($q_{\text{std}} = 50$) | 41 | 48 | 81 | 0.0 |
| nas ($q_{\text{std}} = 20$) | 42 | 48 | 81 | 0.0 |

Plots from a sample trajectory showing the estimation error and covariance bounds for relative velocity and target acceleration using the nca process noise model are shown in Fig. 7.

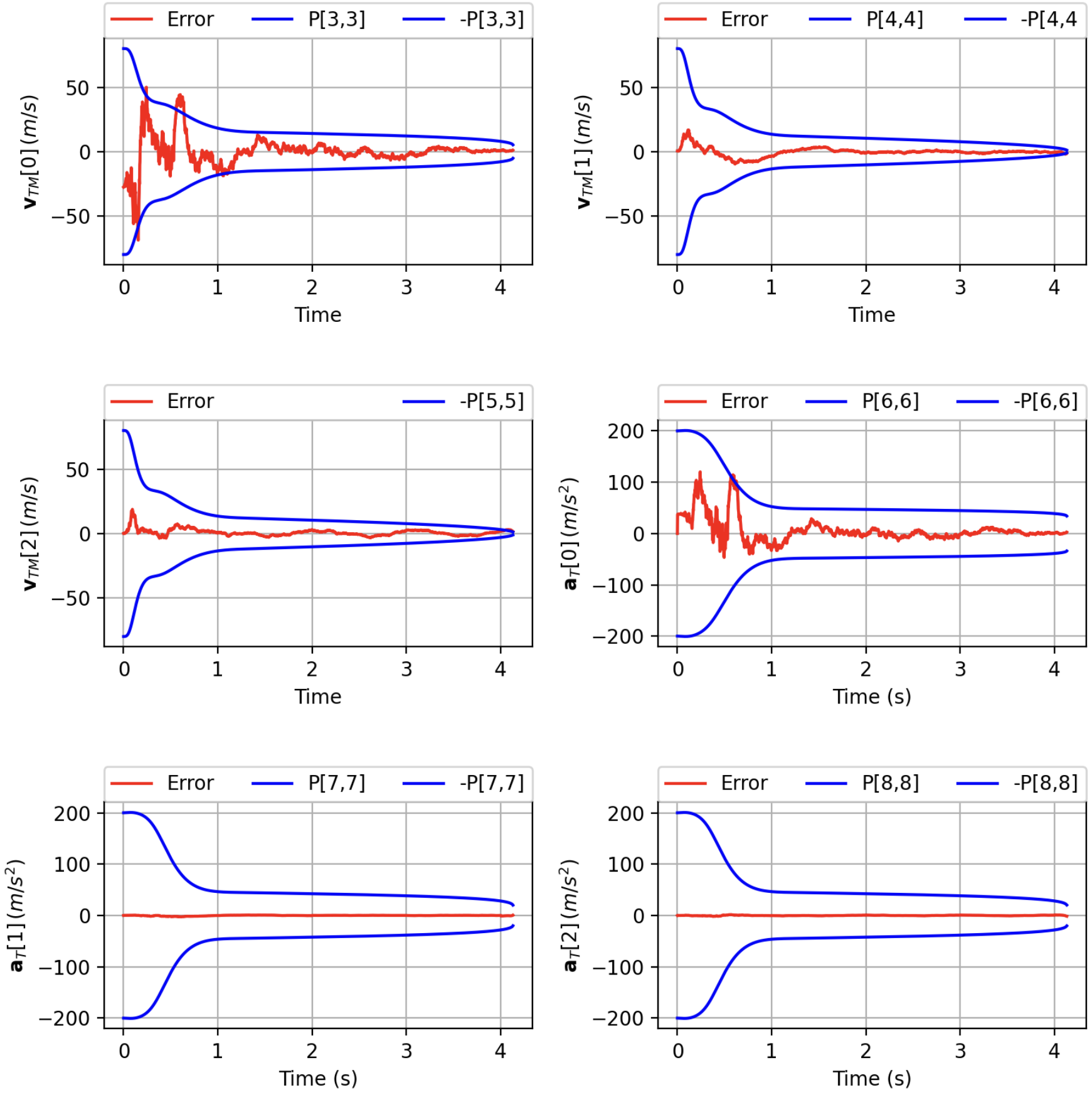

**Fig. 7 UKF Tracking Error**

### B. Interacting Multiple Models

We use the IMM filter [11, 12] as our state-of-the-art adaptive filtering benchmark. The UKF benchmark described in Section III.A assumes that the modeled target dynamics corresponded to the true target dynamics, albeit with different process noise formulations accounting for any mismatches. However, for complex target maneuvers, the filter needs to be able to accurately adapt to rapid changes in the dynamics that cannot be otherwise simply absorbed into the filter as noise. Our interacting multiple-model (IMM) estimator employs seven unscented Kalman Filters, each representing a different constant-jerk hypothesis for the target: zero jerk and positive or negative jerk along each of the three coordinate axes, as shown in Eq. (23).

$$\mathbb{U} = \left\{ \begin{bmatrix} -\mathrm{j}_{\max} \\ 0 \\ 0 \end{bmatrix}, \begin{bmatrix} \mathrm{j}_{\max} \\ 0 \\ 0 \end{bmatrix}, \begin{bmatrix} 0 \\ -\mathrm{j}_{\max} \\ 0 \end{bmatrix}, \begin{bmatrix} 0 \\ \mathrm{j}_{\max} \\ 0 \end{bmatrix}, \begin{bmatrix} 0 \\ 0 \\ -\mathrm{j}_{\max} \end{bmatrix}, \begin{bmatrix} 0 \\ 0 \\ \mathrm{j}_{\max} \end{bmatrix}, \begin{bmatrix} 0 \\ 0 \\ 0 \end{bmatrix} \right\} \tag{23}$$

The initial mode probability vector, $\boldsymbol{\mu}_0$, specifies the prior probability of each model, while the model transition probability matrix, $\mathbf{M}$, defines the probability $M_{ij}$ of transitioning from model $i$ to model $j$ at each filter update step. At each step, the predicted mode probabilities are computed as $\bar{\mu}_k^{(j)} = \sum_{i=1}^{7} M_{ij}\mu_{k-1}^{(i)}$ and the corresponding mixing probabilities $\mu_{k-1}^{(i|j)} = \dfrac{M_{ij}\mu_{k-1}^{(i)}}{\bar{\mu}_k^{(j)}}$ are used to combine the previous model-conditioned state estimates and covariances, providing the initial conditions for each UKF prediction. Each UKF then independently propagates and updates its state estimate according to its assumed jerk model. The mode probabilities are updated using the measurement likelihoods, $\mathcal{L}_k^{(j)}$ and normalized such that $\mu_k^{(j)} \propto \mathcal{L}_k^{(j)} \bar{\mu}_k^{(j)}$. Finally, individual state estimates and covariances are combined using the updated mode probabilities to obtain the overall IMM state estimate and covariance. In our implementation, we use $\mathrm{j}_{\max} = 300$ m/s$^3$, and we filter the likelihoods from each UKF by passing them through a single pole low pass filter with a 20 Hz bandwidth to suppress noise. The initial mode probabilities are $\mu_0 = \begin{bmatrix} 0.25 & 0.125 & 0.125 & 0.125 & 0.125 & 0.125 & 0.125 \end{bmatrix}$ and the fixed model transition probability matrix $\mathbf{M}$ matrix is shown in Eq. (24). These parameters were selected to maximize performance and selected using a coarse grid search.

$$\mathbf{M} = \begin{bmatrix} 9.83\times10^{-1} & 2.75\times10^{-3} & 2.75\times10^{-3} & 2.75\times10^{-3} & 2.75\times10^{-3} & 2.75\times10^{-3} & 2.75\times10^{-3} \\ 1.81\times10^{-3} & 9.91\times10^{-1} & 4.52\times10^{-3} & 6.79\times10^{-4} & 6.79\times10^{-4} & 6.79\times10^{-4} & 6.79\times10^{-4} \\ 1.81\times10^{-3} & 4.52\times10^{-3} & 9.91\times10^{-1} & 6.79\times10^{-4} & 6.79\times10^{-4} & 6.79\times10^{-4} & 6.79\times10^{-4} \\ 1.81\times10^{-3} & 6.79\times10^{-4} & 6.79\times10^{-4} & 9.91\times10^{-1} & 4.52\times10^{-3} & 6.79\times10^{-4} & 6.79\times10^{-4} \\ 1.81\times10^{-3} & 6.79\times10^{-4} & 6.79\times10^{-4} & 4.52\times10^{-3} & 9.91\times10^{-1} & 6.79\times10^{-4} & 6.79\times10^{-4} \\ 1.81\times10^{-3} & 6.79\times10^{-4} & 6.79\times10^{-4} & 6.79\times10^{-4} & 6.79\times10^{-4} & 9.91\times10^{-1} & 4.52\times10^{-3} \\ 1.81\times10^{-3} & 6.79\times10^{-4} & 6.79\times10^{-4} & 6.79\times10^{-4} & 6.79\times10^{-4} & 4.52\times10^{-3} & 9.91\times10^{-1} \end{bmatrix} \tag{24}$$

The performance of our IMM benchmark is shown in Table 3 along with that of the best performing individual UKF. We see that the IMM gives a significant performance improvement and a reduction in target acceleration estimation error.

**Table 3 Benchmark Performance**

| Filter | State Dimension | Miss < 25cm (%) | RMS $\lVert\mathbf{a}_{\mathrm{M}}\rVert$ (m/s$^2$) | RMS $\lVert\mathbf{a}_{\mathrm{T}}^{\mathrm{E}}\rVert$ (m/s$^2$) | Vio (%) |
|---|---|---|---|---|---|
| nas ($q_{\mathrm{std}} = 20$) | 9 | 42 | 48 | 81 | 0.0 |
| IMM (ncj, $q_{\mathrm{std}} = 50$) | 9 | 49 | 47 | 41 | 0.0 |

# IV. Meta-RL Optimization Framework

Note that in this section, unless a variable has the "B" (body frame) superscript, it is assumed to be in the LOS stabilized ("S") reference frame.

## A. Overview of Reinforcement Learning

In the reinforcement learning framework, an agent learns through episodic interaction with an environment how to successfully complete a task using a policy $\pi$ that maps observations $\mathbf{o}$ to actions $\mathbf{u}$, i.e., $\mathbf{u} = \pi(\mathbf{o})$. In actor-critic RL algorithms, policy optimization makes use of a value function $v = V^{\pi}(\mathbf{x})$ that estimates the estimated value of being in state $\mathbf{x}$ while following policy $\pi$. Note that the value function can take either ground truth states $\mathbf{x}$, observations $\mathbf{o}$, or some combination of both, as only the policy is deployed. The environment initializes an episode by randomly generating a ground truth state $\mathbf{x}$, mapping this state to an observation, and passing the observation to the agent. The agent uses this observation to generate an action that is sent to the environment; the environment then uses the action and the current ground truth state to generate the next state and a scalar reward signal $r(\mathbf{x}, \mathbf{u})$. The reward and the observation corresponding to the next state are then passed to the agent. The process repeats until the environment terminates the episode, with the termination signaled to the agent via a done signal. In this work, and episode terminates when the closing velocity $v_c$ turns negative, or a path constraint is violated. Trajectories collected over a set of episodes (referred to as rollouts) are collected during interaction between the agent and environment, and used to update the policy and value functions. The interface between agent and environment is depicted in Fig. 8, where the environment instantiates the simulator and the system model shown in Fig. 1.

Meta-RL differs from generic reinforcement learning in that the agent learns over an ensemble of environments. For example, each environment can contain a unique aerodynamic model, target maneuver model, actuator model, and other peripheral components. Similar to [30], we implement both the policy and value function as a deep recurrent neural network, with each network having a hidden state $\mathbf{h}$ that is a function of the history of observations $\mathbf{o}$. Thus, the policy and value function become $\mathbf{u} = \pi_\theta(\mathbf{o}, \mathbf{h})$ and $v = V^\pi(\mathbf{x}, \mathbf{h})$. We optimize the policy using a customized version of the proximal policy optimization algorithm (PPO) [31]. Maximizing the PPO objective function requires learning hidden layer parameters $\boldsymbol{\theta}_\mathrm{h}$ that result in $\mathbf{h}$ evolving in response to the history of $\mathbf{o}$ and $\mathbf{u}$ in a manner that facilitates fast adaptation to environments encountered during optimization, as well as novel environments outside of the training distribution. The optimized policy then adapts real-time to conditions experienced during deployment. Importantly, the network parameters remain fixed during deployment with adaptation occurring through the evolution of $\mathbf{h}$.

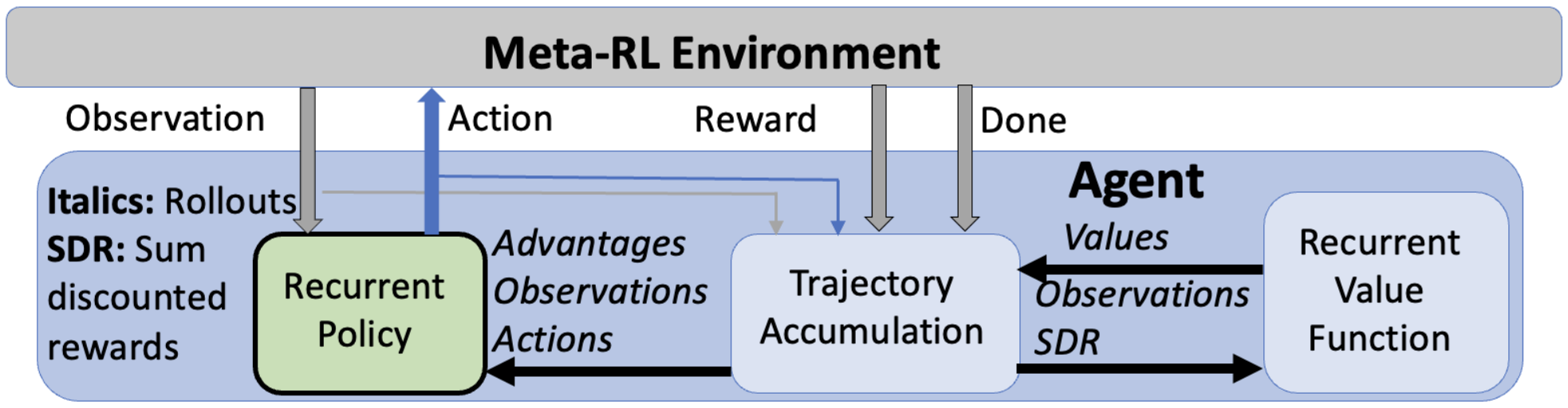

**Fig. 8 Environment-Agent Interface**

The proximal policy optimization (PPO) [31] algorithm used in this work is a policy gradient algorithm which has demonstrated state-of-the-art performance for many reinforcement learning benchmark problems. PPO approximates the Trust Region Policy Optimization method [32] by accounting for the policy adjustment constraint with a clipped objective function. The objective function used with PPO can be expressed in terms of the probability ratio $p_k(\boldsymbol{\theta})$ given by,

$$p_k(\boldsymbol{\theta}) = \frac{\pi_{\boldsymbol{\theta}}(\mathbf{u}_k|\mathbf{o}_k)}{\pi_{\boldsymbol{\theta}_\mathrm{old}}(\mathbf{u}_k|\mathbf{o}_k)} \tag{25}$$

The PPO objective function is shown in Equations (26a) through (26c). The general idea is to create two surrogate objectives, the first being the probability ratio $p_k(\boldsymbol{\theta})$ multiplied by the advantages $A^\pi_\mathbf{w}(\mathbf{o}_k, \mathbf{u}_k)$ (see Eq. (27)), and the second a clipped (using clipping parameter $\epsilon$) version of $p_k(\boldsymbol{\theta})$ multiplied by $A^\pi_\mathbf{w}(\mathbf{o}_k, \mathbf{u}_k)$. The objective to be maximized $J(\boldsymbol{\theta})$ is then the expectation under the trajectories induced by the policy of the lesser of these two surrogate objectives.

$$\mathrm{obj1} = p_k(\boldsymbol{\theta}) A^\pi_\mathbf{w}(\mathbf{o}_k, \mathbf{u}_k) \tag{26a}$$

$$\mathrm{obj2} = \mathrm{clip}(p_k(\boldsymbol{\theta}) A^\pi_\mathbf{w}(\mathbf{o}_k, \mathbf{u}_k), 1-\epsilon, 1+\epsilon) \tag{26b}$$

$$J(\boldsymbol{\theta}) = \mathbb{E}_{p(\boldsymbol{\tau})}[\min(\mathrm{obj1}, \mathrm{obj2})] \tag{26c}$$

This clipped objective function has been shown to maintain a bounded Kullback-Leibler (KL) divergence [33] with respect to the policy distributions between updates, which aids convergence by ensuring that the policy does not change drastically between updates. Our implementation of PPO uses an approximation to the advantage function that is the difference between the empirical return and a state value function baseline, as shown in Equation 27, where $\gamma$ is a discount rate and $r$ the reward function, described in Section IV.B.

$$A^\pi_\mathbf{w}(\mathbf{x}_k, \mathbf{u}_k) = \left[\sum_{\ell=k}^{T} \gamma^{\ell-k} r(\mathbf{x}_\ell, \mathbf{u}_\ell)\right] - V^\pi_\mathbf{w}(\mathbf{x}_k) \tag{27}$$

Here the value function $V^\pi_\mathbf{w}$ is learned using the cost function given by

$$L(\mathbf{w}) = \frac{1}{2M} \sum_{i=1}^{M} \left( V^\pi_\mathbf{w}(\mathbf{x}^i_k) - \left[\sum_{\ell=k}^{T} \gamma^{\ell-k} r(\mathbf{u}^i_\ell, \mathbf{x}^i_\ell)\right]\right)^2 \tag{28}$$

In practice, policy gradient algorithms update the policy using a batch of trajectories (roll-outs) collected by interaction with the environment. Each trajectory is associated with a single episode, with a sample from a trajectory collected at step $k$ consisting of observation $\mathbf{o}_k$, action $\mathbf{u}_k$, and reward $r_k(\mathbf{o}_k, \mathbf{u}_k)$. Finally, gradient ascent is performed on $\boldsymbol{\theta}$ and gradient descent on $\mathbf{w}$ and update equations are given by

$$\mathbf{w}^{+} = \mathbf{w}^{-} - \beta_{\mathbf{w}} \nabla_{\mathbf{w}} \, L(\mathbf{w})|_{\mathbf{w}=\mathbf{w}^{-}} \tag{29}$$

$$\boldsymbol{\theta}^{+} = \boldsymbol{\theta}^{-} + \beta_{\boldsymbol{\theta}} \, \nabla_{\boldsymbol{\theta}} J(\boldsymbol{\theta})|_{\boldsymbol{\theta}=\boldsymbol{\theta}^{-}} \tag{30}$$

where $\beta_{\mathbf{w}}$ and $\beta_{\boldsymbol{\theta}}$ are the learning rates for the value function, $V^{\pi}_{\mathbf{w}}(\mathbf{o}_k)$, and policy, $\pi_{\boldsymbol{\theta}}(\mathbf{u}_k|\mathbf{o}_k)$, respectively.

In our implementation of PPO, we adaptively scale the observations and servo both $\epsilon$ and the learning rate to target a KL divergence of 0.001.

### B. Meta-RL Problem Formulations

The observations $\mathbf{o}$ are normalized to have zero mean and unit variance. The primary considerations for constructing the policy observation $\mathbf{o}$ is that the observations need to be available from sensor outputs, and that these observations are useful in generating the action $\mathbf{u}$ in a way that maximizes the advantages $A^{\pi}_{\mathbf{w}}$. The observation space for the meta-RL jerk estimation policy is shown in Eqs. (31), where $\mathbf{x}_{\mathrm{f}}$ is the current filter state, $\mathbf{y}_{\mathrm{f}}$ the filter innovation (difference between predicted and observed measurement), and $\mathbf{x}_{\mathrm{post}} - \mathbf{x}_{\mathrm{prior}}$ is the forward update difference (the difference between the updated state incorporating the measurement and the prior state output by the prediction function). All observations are in the quasi-inertial stabilized reference frame as described in II.C.5. For some intuition as to the policy observation space construction, $\mathbf{x}_{\mathrm{f}}$ shows the policy how its jerk estimate impacts the estimate, and the forward update difference and innovation give the policy access to some of the intermediate filter computations.

$$\mathbf{o}_{\pi} = \begin{bmatrix} \mathbf{x}_{\mathrm{f}} & \mathbf{y}_{\mathrm{f}} & \mathbf{x}_{\mathrm{post}} - \mathbf{x}_{\mathrm{prior}} \end{bmatrix} \tag{31}$$

We use a slightly different observation space for the value function during optimization, replacing $\mathbf{x}_{\mathrm{f}}$ with the state estimation error $\mathbf{e}_{\mathrm{f}}$, which is the difference between the predicted and ground truth state. This is allowable because the value function is only used for optimization and is not deployed along with the policy. The value function observation space is shown in Eq. (32).

$$\mathbf{o}_{v} = \begin{bmatrix} \mathbf{e}_{\mathrm{f}} & \mathbf{y}_{\mathrm{f}} & \mathbf{x}_{\mathrm{post}} - \mathbf{x}_{\mathrm{prior}} \end{bmatrix} \tag{32}$$

The output of the policy $\mathbf{u} = \pi(\mathbf{o}) \in \mathbb{R}^3$ is scaled through multiplication by the largest expected target jerk, which we estimate to be 20000 m/s$^3$ which corresponds to a change in target acceleration of 40 m/s$^2$ between samples, with the state estimator running at 500 Hz. This estimate then replaces the zeros vector in Eq. (20c) in Section III.

The reward function for the jerk estimator is shown below in Equations (33a) through (33d). $r_{\mathrm{shaping1}}$ is a shaping reward given at each step in an episode. These shaping rewards take the form of a Gaussian-like function of the norm of the state estimation error $\mathbf{e}_{\mathrm{f}}$ in log space as shown in Eq. (33a). We transform to log space as this error can be very high during the early part of optimization. Similarly, $r_{\mathrm{shaping2}}$ uses $\mathbf{e}_{\mathrm{f}}$ in linear space, with this reward component only becoming significant later in the optimization process. Finally, $r_{\mathrm{ctrl}}$ is a control effort penalty, where $\mathbf{u}$ is the unscaled policy action. We use $\alpha = 1$, $\sigma_{\alpha} = 2$, $\beta = 1$, $\sigma_{\beta} = 20$, $\kappa = 0.01$, $\sigma_{\kappa} = 0.10$, and $\delta = -0.01$. $a_{\mathrm{max}}$ is the interceptors maximum allowed acceleration at sea level. We use a discount rate (see Section IV.A of 0.95.

$$r_{\mathrm{shaping1}} = \alpha \exp\left(\frac{(\log(\|\mathbf{e}_{\mathrm{f}}\| + 1))^2}{\sigma_{\alpha}^2}\right) \tag{33a}$$

$$r_{\mathrm{shaping2}} = \beta \exp\left(\frac{(\|\mathbf{e}_{\mathrm{f}})\|)^2}{\sigma_{\beta}^2}\right) \tag{33b}$$

$$r_{\mathrm{ctrl}} = \delta\|\mathbf{u}\| \tag{33c}$$

$$r = r_{\mathrm{shaping1}} + r_{\mathrm{shaping2}} + r_{\mathrm{ctrl}} \tag{33d}$$

The policy and value functions are implemented using four layer neural networks with tanh activations on each hidden layer. Layer 2 for the policy and value function is a recurrent layer implemented using gated recurrent units [34]. The network architectures are as shown in Table 4, where $n_{\mathrm{hi}}$ is the number of units in layer $i$, obs_dim is the observation dimension, and act_dim is the action dimension. The policy and value functions are periodically updated during optimization after accumulating trajectory rollouts of 30 simulated episodes.

Table 4 **Policy and Value Function network architecture**

| | Policy Network | | Value Network | |
|---|---|---|---|---|
| Layer | # units | activation | # units | activation |
| hidden 1 | $10*\text{obs_dim}$ | tanh | $10*\text{obs_dim}$ | tanh |
| hidden 2 | $\sqrt{n_{h1} * n_{h3}}$ | tanh | $\sqrt{n_{h1} * n_{h3}}$ | tanh |
| hidden 3 | $10 * \text{act_dim}$ | tanh | 5 | tanh |
| output | act_dim | linear | 1 | linear |

# V. Experiments

## A. Meta-RL FCS Optimization

The policy is optimized over the range of initial conditions given in Table 1, with the target maneuver distribution a 50% probability of weaving and a 50% probability of jinking. Step maneuvers are held out to test policy generalization. The reward learning curve for the initial optimization is shown in Fig. 9, where "Mean" is the mean reward over a set of rollouts. The reward history is slightly noisy due to the fact that the exploration variance does not fall to zero during the optimization (it would be destabilizing to the PPO algorithm if it did fall to zero). Note that for testing, the exploration variance is disabled, and we return the mean value of the policy distribution.

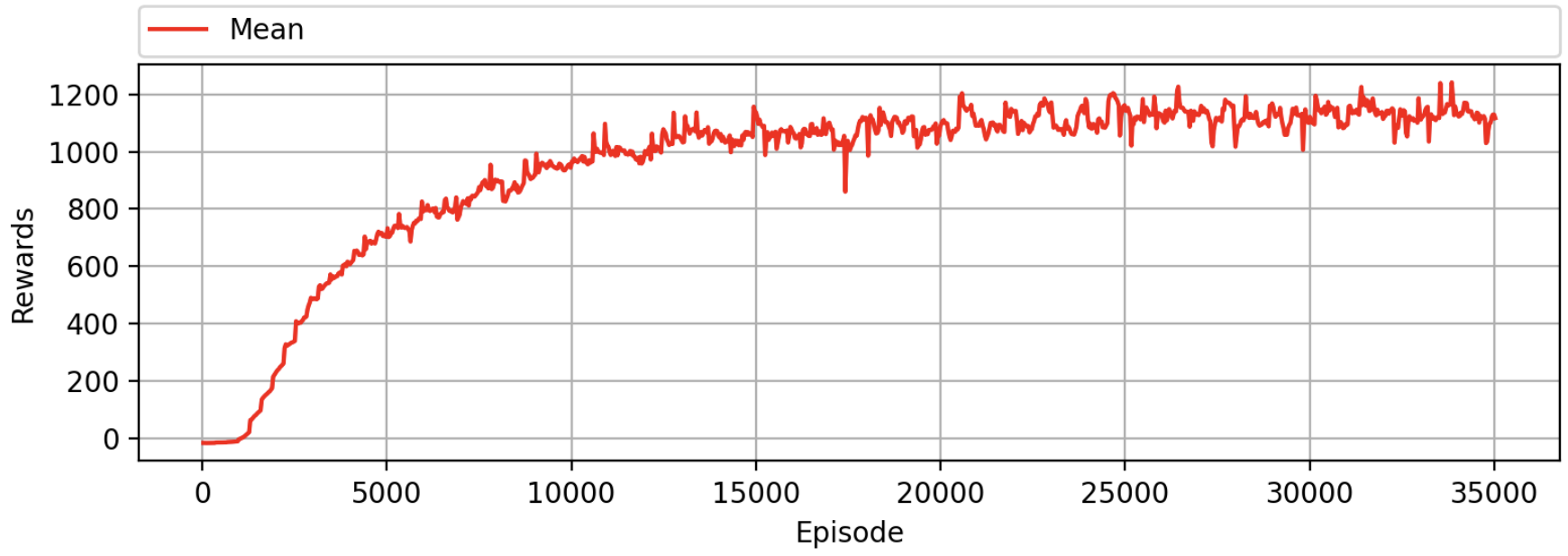

**Fig. 9 Optimization Reward History**

## B. Filter Relative Performance

Relative performance is measured by running 5000 simulations for each case in the following tables. Table 5 compares the interceptor performance over various maneuver types using different state estimation implementations. Note that the "J+W" case has the target executing a jink maneuver in 50% of the episodes and a weave maneuver in the other 50%. "GT" indicates performance without the UKF in the loop, using the noisy $\mathbf{r}_{\text{TM}}$ obtained from range, elevation, and azimuth measurements, and ground truth target relative velocity. The "UKF" case uses the nine-state UKF from Section III.A using the Singer process noise model (nas). The "IMM" case uses the IMM implementation developed in Section III.B with each UKF using the near-constant jerk (ncj) process noise model. The "UKF-RL" case uses the meta-RL policy to estimate target jerk, with the estimate fed into the propagation model as described in Section III. Here the UKF uses a near-constant acceleration (nca) process noise model, which we found provides the best performance when combined with jerk estimation. "IR" is the interceptor ratio, which denotes the number of interceptors using the benchmark UKF that are required to match the accuracy of an interceptor using the meta-RL augmented UKF, according to the equation $\text{IR} = \dfrac{\log m_1}{\log m_2}$, where $m_1$ and $m_2$ are the probability of a miss distance greater than 25 cm for the meta-RL and benchmark UKF respectively.

First, we see that regardless of the state estimation method, there is a large reduction in performance with the state estimator in the loop. This is expected due to the combination of realistic noise sources (many of which are not Gaussian and therefore violate the UKF process and measurement noise assumptions) and a highly agile target executing challenging maneuvers. The most challenging maneuver is the weaving maneuver, followed by the jinking maneuver.

**Table 5 Performance Comparison**

| Maneuver | UKF | Miss < 25cm (%) | RMS $\|\mathbf{a}_M\|$ (m/s$^2$) | RMS $\|\mathbf{a}_T^E\|$ (m/s$^2$) | IR | Vio (%) |
|---|---|---|---|---|---|---|
| Weave | **GT** | 91 | 43 | - | - | 0.0 |
| Weave | **UKF** | 31 | 48 | 85 | 2.5 | 0.0 |
| Weave | **IMM** | 42 | 47 | 40 | 1.6 | 0.1 |
| Weave | **UKF-RL** | 60 | 45 | 34 | - | 0.0 |
| Jink | **GT** | 90 | 45 | - | - | 0.0 |
| Jink | **UKF** | 43 | 48 | 81 | 1.8 | 0.0 |
| Jink | **IMM** | 52 | 48 | 40 | 1.4 | 0.0 |
| Jink | **UKF-RL** | 64 | 47 | 35 | - | 0.0 |
| J+W | **GT** | 91 | 44 | - | - | 0.0 |
| J+W | **UKF** | 40 | 47 | 82 | 1.9 | 0.0 |
| J+W | **IMM** | 47 | 48 | 41 | 1.5 | 0.0 |
| J+W | **UKF-RL** | 62 | 46 | 35 | - | 0.0 |

Comparing the different filter types, IMM gives a moderate performance improvement over the UKF, with the UKF-RL filter giving a similar moderate improvement over the IMM. Although the performance improvement is modest, it is significant when considering the number of interceptors required for the UKF or IMM to match the interception probability of the UKF-RL filter.

Table 6 illustrates the impact of noise and target acceleration on filter performance. The "No Noise" case has IMU noise and bias set to zero, radar receiver angle and range noise set to zero, and no radome refraction. The lack of IMU noise allows for a near perfect recovery of the LOS in the stabilized reference frame (see Section II.C.5), with imperfect LOS recovery being a significant source of non-linear noise. The lack of radome refraction removes another source of non-linear process and measurement noise. The "TACC=0" case has the same noise parameters as in Table 5, but with target evasive acceleration set to zero. In this case the interceptor will still perceive the component of target acceleration due to drag perpendicular to the LOS as a maneuver. The "Ideal" case has no noise or target acceleration. For convenience, the "Realistic" case repeats the last rows of Table 5. With respect to the "Ideal" case, it appears that although noise sources lead to a significant decrease in performance, the actively evading target results in the largest performance decrease.

**Table 6 Impact of Noise and Target Acceleration, J+W Maneuver**

| Case | UKF | Miss < 25cm (%) | RMS $\|\mathbf{a}_M\|$ (m/s$^2$) | RMS $\|\mathbf{a}_T^E\|$ (m/s$^2$) | IR | Vio (%) |
|---|---|---|---|---|---|---|
| Ideal | **UKF** | 100 | 42 | 31 | 1.0 | 0.0 |
| Ideal | **IMM** | 100 | 42 | 23 | 1.0 | 0.0 |
| Ideal | **UKF-RL** | 100 | 42 | 25 | - | 0.0 |
| TACC=0 | **UKF** | 81 | 43 | 84 | 1,2 | 0.0 |
| TACC=0 | **IMM** | 71 | 44 | 40 | 1.6 | 0.0 |
| TACC=0 | **UKF-RL** | 86 | 43 | 33 | - | 0.0 |
| No Noise | **UKF** | 53 | 48 | 32 | 2.1 | 0.0 |
| No Noise | **IMM** | 69 | 46 | 24 | 1.3 | 0.0 |
| No Noise | **UKF-RL** | 79 | 45 | 26 | - | 0.0 |
| Realistic | **UKF** | 40 | 47 | 82 | 1.9 | 0.0 |
| Realistic | **IMM** | 47 | 48 | 41 | 1.5 | 0.0 |
| Realistic | **UKF-RL** | 62 | 46 | 35 | - | 0.0 |

Table 7 gives interceptor and target statistics computed over 5000 trajectories, where we see that the acceleration of the interceptor and target are quite close at the end of the engagement, making this a highly agile target. Since the target starts at a much higher altitude than the interceptor, the lower dynamic pressure gives the target less maneuverability earlier in the engagement.

**Table 7 High Altitude Engagement Statistics**

| | |
|---|---|
| Cumulative RMS $\|\|\mathbf{a}_{\mathrm{M}}^{\mathrm{B}}[y,z]\|\|$ (m/s$^2$) | 40 |
| Cumulative RMS $\|\|\mathbf{a}_{\mathrm{T}}\|\| \perp \lambda_{\mathrm{TM}}$ (m/s$^2$) | 12 |
| Terminal RMS $\|\|\mathbf{a}_{\mathrm{M}}^{\mathrm{B}}[y,z]\|\|$ (m/s$^2$) | 58 |
| Terminal RMS $\|\|\mathbf{a}_{\mathrm{T}}\|\| \perp \lambda_{\mathrm{TM}}$ (m/s$^2$) | 51 |
| Maximum Terminal Altitude $h$ (m) | 30347 |

## C. Policy Generalization Testing

We demonstrate the generalization capability of the jerk estimation policy by increasing the variation of initial range to target and initial target speed, as shown in Table 8. The expanded initial condition simulations ("Ext-IC") use a combination of weaving and jinking maneuvers, each with 50% probability, and are consequently comparable to rows 14 and 16 in Table 9. We also test increased initial state estimation uncertainty ("Ext-U") by doubling this from the values used in Section III. This differs from the "Ext-IC" case in that it is the uncertainty in the filter's initial state estimate that is increased. Moreover, we test against three target maneuver types not seen during optimization, a step maneuver, a vertical S maneuver [35] (vert-S), and frequency modulated weave (FM-weave). The vertical S maneuver is identical to the weave maneuver except that the target acceleration instantaneously transitions between minimum and maximum acceleration at the weave maneuver zero acceleration crossings. In our simulator the vertical S differs from the jinking maneuver in that the jinking maneuver does not switch sign with a fixed period, and the jinking is independent for each axis. The frequency modulated weave has a random variation in the weave frequency over each engagement. This would be a relatively simple maneuver to execute for a maneuverable ballistic re-entry vehicle that is already spiraling with the fin deflections used to modulate the spiraling frequency. Example target acceleration trajectories for the FM-weave maneuver are illustrated in Fig. 10

**Table 8 Simulation Expanded Initial Conditions)**

| Parameters Drawn Uniformly | Min | Max |
|---|---|---|
| Initial Range to Targeted Asset $\|\|\mathbf{r}_{\mathrm{AT}}\|\|$ (m) | 15000 | 30000 |
| Target Velocity Magnitude $\|\|\mathbf{v}_{\mathrm{T}}\|\|$ (m/s) | 2500 | 4500 |

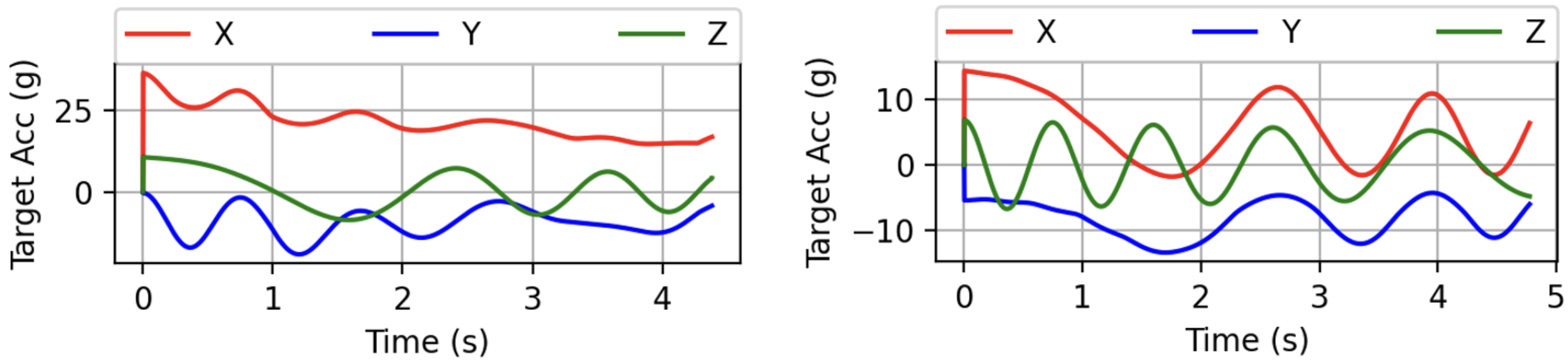

**Fig. 10 Frequency Modulated Weave Sample Trajectories**

From Table 9 we see that all three filters are robust to an increase in both the range of initial conditions and uncertainty in the initial state estimate. The vertical-S maneuver is challenging for all three filters. The frequency modulated weave proves challenging for the UKF and IMM, whereas there is only a small performance drop for the UKF-RL filter. In contrast, the vertical-S maneuver is challenging for all three filters, but more so for the UKF-RL if we look at relative drop in performance from a jink maneuver. Interestingly, both the IMM and UKF-RL slightly under perform the UKF for the step maneuver. The IMM performance could likely have been improved by optimizing the IMM hyperparameters with a step maneuver included in the target maneuver distribution, at a potential decrease in performance for the more challenging target maneuvers, and similarly for the UKF-RL. The key takeaway is that the UKF-RL jerk estimation policy generalizes well to both initial conditions and target maneuvers outside of the optimization distribution.

**Table 9 Target Maneuver Generalization Cases**

| Maneuver | UKF | Miss < 25cm (%) | RMS $\|\mathbf{a}_M\|$ (m/s$^2$) | RMS $\|\mathbf{a}_T^E\|$ (m/s$^2$) | IR | Vio (%) |
|---|---|---|---|---|---|---|
| J+W, Ext-IC | **UKF** | 50 | 54 | 84 | 1.5 | 0.1 |
| J+W, Ext-IC | **IMM** | 52 | 46 | 40 | 1.4 | 0.1 |
| J+W, Ext-IC | **UKF-RL-J** | 65 | 47 | 35 | - | 0.1 |
| J+W, Ext-U | **UKF** | 41 | 50 | 68 | 1.8 | 0.5 |
| J+W, Ext-U | **IMM** | 47 | 49 | 41 | 1.5 | 0.6 |
| J+W, Ext-U | **UKF-RL-J** | 61 | 49 | 32 | - | 0.1 |
| Step | **UKF** | 83 | 47 | 84 | 0.7 | 0.0 |
| Step | **IMM** | 72 | 46 | 40 | 1.0 | 0.0 |
| Step | **UKF-RL** | 72 | 48 | 34 | - | 0.0 |
| vert-S | **UKF** | 26 | 50 | 84 | 2.0 | 0.0 |
| vert-S | **IMM** | 31 | 49 | 41 | 1.7 | 0.0 |
| vert-S | **UKF-RL** | 46 | 47 | 34 | - | 0.0 |
| FM-weave | **UKF** | 23 | 49 | 85 | 3.3 | 0.0 |
| FM-weave | **IMM** | 31 | 48 | 41 | 2.3 | 0.0 |
| FM-weave | **UKF-RL** | 58 | 46 | 34 | - | 0.0 |

# VI. Conclusion

We demonstrated the use of reinforcement learning to optimize a target jerk estimation policy integrated with a standard nine-state unscented Kalman filter (UKF) using a near-constant acceleration process noise model, injecting the target jerk estimate into the filter's propagation model. Moreover, after testing multiple UKF architectures, two benchmarks were selected, a UKF with a Singer process noise model and an implementation of interacting multiple models (IMM) with each model implementing a different jerk hypothesis and using a near-constant jerk process noise model. All three filters were optimized or tuned in a high-fidelity 6-DOF simulator modeling a challenging intercept application. The filters were then tested in the same environment, and we explored the impact of target maneuver type and noise on filter performance. Importantly, we tested the filter generalization performance by expanding the range of initial conditions, initial filter uncertainty, and target maneuver types. We found that the IMM outperformed (both accuracy and control effort) the benchmark UKF by a significant margin, and that the meta-RL augmented UKF (UKF-RL) outperformed the IMM by a similar margin. One possible factor behind the improved performance using the UKF-RL is that the IMM selects the model using a single scalar observation, the filtered likelihood, whereas the UKF-RL policy has a more extensive observation space see (Section IV.B). Indeed, if we reduce the UKF-RL observation space to just the innovation, we find a reduction in performance. Other factors include the finer granularity of target jerk estimates (continuous versus seven models), the fixed transition probability matrix implemented by the IMM, and the recurrent neural network implemented in the UKF-RL. In contrast to recent methods that implement the entire filter as one or more neural networks, our approach augments a standard UKF, simplifying the process of flight certification. The performance improvement demonstrated in this work suggests that the UKF-RL filter approach has the potential to significantly increase interceptor effectiveness.